\documentclass{ws-p8-50x6-00}

\begin{document}

\title{Prospects for HERMES Run II}

\author{Wolf-Dieter Nowak}

\address{on behalf of the HERMES collaboration} 

\address{DESY Zeuthen, D-15738 Zeuthen, Germany \\
E-mail: Wolf-Dieter.Nowak@desy.de}  


\maketitle

\abstracts{Data taking for Run II of the HERMES experiment will 
start in late 2001 with three main physics objectives for the next
4-5 years: a measurement of transversity distributions,
an improved measurement of helicity distributions, and
measurements of exclusive reactions to access Generalized Parton
Distributions.}

\section{Introduction}
The HERMES experiment~\cite{spectr} at DESY Hamburg, Germany, combines 
a forward spectrometer with an internal gas target in the 27.5~GeV 
lepton ring of the HERA electron(positron)-proton collider. In between a
pair of spin rotators enclosing the experiment the circulating 
self-polarized electrons (or positrons) get their transverse polarization 
changed into a longitudinal one; both 
polarizations are permanently monitored. The open-ended
storage cell allows for high target gas densities: $10^{14}$ 
nucl/cm$^2$ are reachable for polarized hydrogen and 
deuterium, while the density of unpolarized gases can be 
up to 100~times higher, limited by beam and spectrometer conditions.

Systems of drift chambers before and after the 1.3 Tm dipole 
spectrometer magnet yield a tracking resolution of $\delta p/p \simeq 1 \%$. 
Together with a set of scintillator hodoscopes the lead-glass 
electromagnetic calorimeter is used for high efficiency
triggering of electrons and positrons. In conjunction with a TRD, a
preshower and a Cerenkov detector it also allows for an efficient 
lepton-hadron separation. The Cerenkov detector was upgraded in 1998 
from a threshold to a RICH detector~\cite{rich}, now delivering
pion/kaon/proton separation over most of the momentum range, 
$2 < p < 16$~GeV. 

Over 40 millions DIS events (after data quality cuts), corresponding to an
integrated luminosity of almost 1 fb$^{-1}$,  were collected
at HERMES in the years 1995-2000. About one third was taken with a 
(longitudinally) polarized target: 2.5 M on helium-3 (1995), 2.5 M on 
hydrogen (1996-97) and 8.5 M on deuterium (1998-2000). Two thirds of
the data originate from several unpolarized target gases, ranging in 
atomic number from hydrogen to krypton.

Originally, the main physics goal of HERMES was the study of polarized 
semi-inclusive deep inelastic scattering (SIDIS). This led to a novel 
feature of the HERMES experiment, as compared to earlier measurements at CERN,  
namely its capability to identify outgoing hadrons in addition 
to the scattered lepton. To reduce the systematic uncertainties of
spin asymmetry measurements, the orientations of target (if polarized) 
and beam spin are regularly flipped, whereby the time scale for
spin flips is seconds for the former and weeks for the latter.

As a first major physics result the helicity distributions of $u$- and 
$d$-quarks were precisely determined over the full accessible range, 
$0.023 < x < 0.6$, while a corresponding separation for sea quarks 
remains an issue. The accuracy attainable at HERMES after the inclusion 
of all already collected and future data are discussed in sect.~3.
The SIDIS data can be analyzed more differentially, especially 
the azimuthal dependence of (beam or target) single-spin 
asymmetries for produced hadrons can be studied. This in principle opens 
access to a variety of `new', i.e. hitherto unmeasured, chirally-odd  
parton distribution and fragmentation functions; prospects are discussed 
in sect.~2. The study of exclusive reactions attained particular theoretical 
interest in the last few years; they appear 
to be a potentially well-suited experimental tool to study Generalized 
(or Skewed) Parton Distributions, GPDs (SPDs). 
At HERMES, (quasi-) exclusive particle production is accessible thanks
to the good hadron and photon detection capabilities of the spectrometer.
Prospects for future HERMES data on exclusive reactions are presented in 
sect.~4.

\section{Measurement of Transversity Distributions}
A quark of given flavour in the nucleon is characterized by three independent
twist-2 quark distributions: the well-measured number density distribution
$q(x)$, the helicity distribution $\Delta q(x)$ measured by now for valence
quarks, and the hitherto unmeasured transversity distribution $\delta q(x)$ 
that characterizes the distribution of the quark's transverse spin in a
transversely polarized nucleon. A complete understanding of the nucleon
structure on the twist-2 level requires precise experimental data for all 
three functions.

Quark couplings with gluons and photons preserve chirality, i.e.
observables have to be chirally even. Hence the chirally-odd transversity 
distributions should enter in combination with other chirally-odd objects in
the corresponding expressions. It turns out that such combinations occur in 
several different SIDIS processes~\cite{KorNowPrag}, but not in inclusive DIS. 

As a representative SIDIS example pion production by an unpolarized beam 
on a transversely polarized target is considered. The following 
{\it $\sin(\phi)$-weighted} single target-spin asymmetry \cite{kotmul97}
provides access to the quark transversity distributions via the Collins effect:
\newpage
\begin{equation}
A_T(x,y,z) \equiv
\frac{\int d \phi^\ell \int d^2P_{h\perp}\,
\frac{\vert P_{h\perp}\vert}{zM_h}
\sin(\phi_s^\ell + \phi_h^\ell)
\,\left(d\sigma^{\uparrow}-d\sigma^{\downarrow}\right)}
{\int d \phi^\ell \int d^2 P_{h\perp} (d\sigma^{\uparrow}+d\sigma^{\downarrow})} .
\label{collasy}
\end{equation}
Here $x$, $y$ and $z$ are the standard SIDIS variables and
$P_{h\perp}$ is the pion's transverse momentum. The azimuthal 
angles are defined in transverse space giving the 
orientation of the hadron plane ($\phi^\ell_h$ = $\phi_h - \phi^\ell$)
or spin vector ($\phi^\ell_s$ = $\phi_s - \phi^\ell$)
with respect to the azimuthal orientation of the lepton plane ($\phi^\ell$).
For the calculation of the expected statistical accuracies the asymmetry 
(\ref{collasy}) can be estimated from
\begin{equation}
\label{asymmetry}
A_T(x,y,z) =
P_T \cdot D_{nn} \cdot
\frac{\sum_q e^2_q \  \delta q(x) \ H_1^{\perp(1)q}(z)}
     {\sum_q e^2_q \  q(x) \ D^q_1(z)} ,
\end{equation}
and $P_T$ is the target polarization,
$D_{nn}(y)$ the virtual photon transverse spin transfer coefficient,
$H_1^{\perp (1) q}(z)$ the T-odd polarized `Collins'  fragmentation 
function and $D^q_1(z)$ the usual unpolarized fragmentation function.

A sine-shaped single target-spin asymmetry with a magnitude of about 
2\% was extracted from 1996-1997 HERMES data collected on a longitudinally polarized 
proton target~\cite{pi-SSA}. It has been interpreted in terms of the 
existence of a non-zero `Collins'
function $H_1^{\perp (1) q}(z)$, in conjunction with 
a non-zero transversity distribution. Using instead a {\it transversely} 
polarized proton target, the size of transverse spin phenomena is expected 
to grow by almost one order of magnitude. 

To calculate prospects for future HERMES running
the assumption of $u$-quark dominance can be used to determine e.g.
the expected asymmetry $A_T^{\pi^+}(x)$~\cite{KNO}.
In this case the asymmetry for a proton target reduces to
\begin{equation}
\label{uasymmetry}
A_T^{\pi^+}(x,y,z) =
P_T \cdot D_{nn} \cdot
\frac{\delta u(x)}{u(x)} \cdot \frac{H_1^{\perp(1)u}(z)}{D^u_1(z)} ,
\end{equation}
and the approach of Ref.~\cite{kotmul97} can be adopted to estimate
$H_1^{\perp(1)u}(z) / D^u_1(z)$.
\begin{figure}[htb]
\centering
\vspace*{-0.4cm}
\epsfxsize=8.5cm
\epsfbox{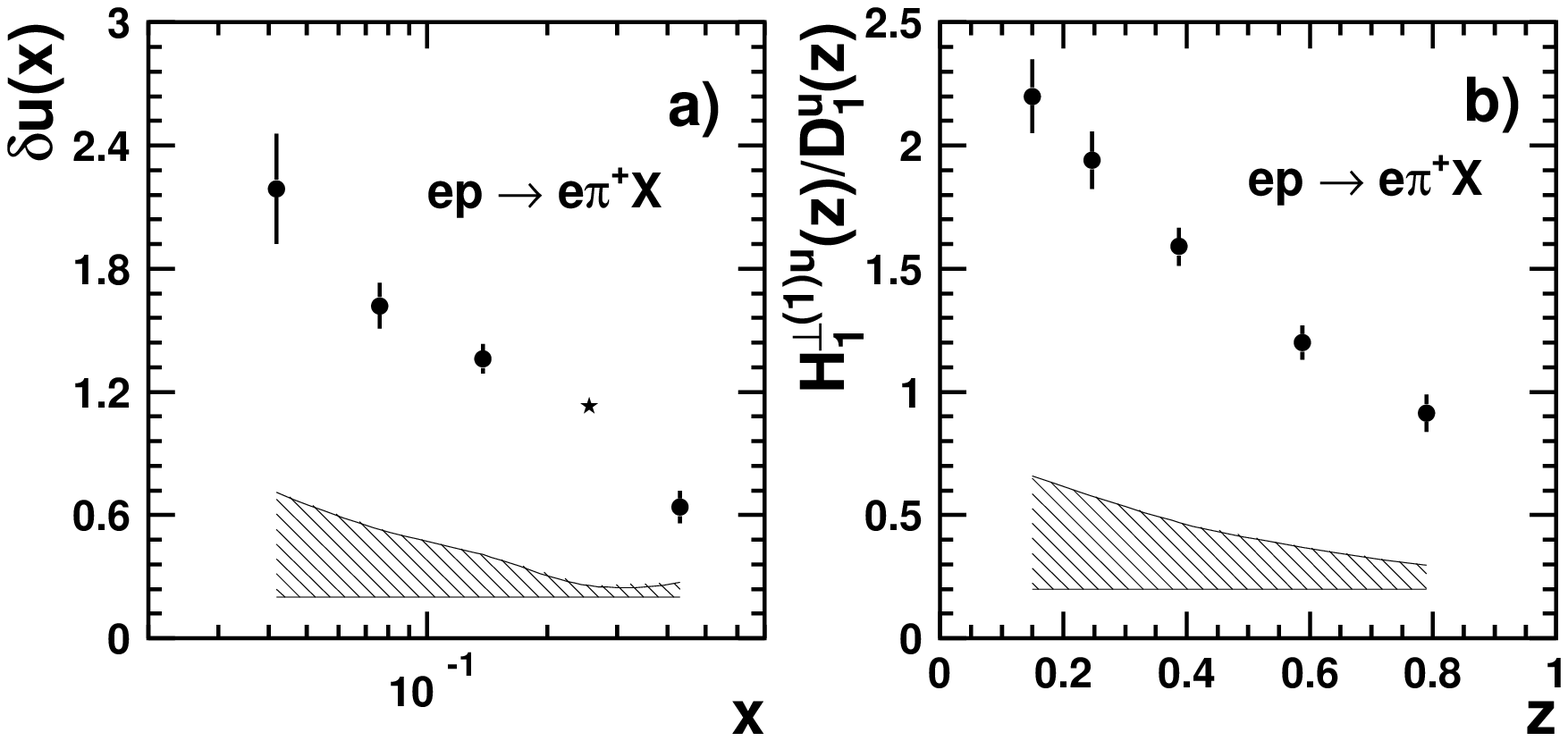} 
\vspace*{-0.7cm}
\caption{Projected statistical precision for transversity measurements on
a transversely polarized proton target: a) transversity distribution 
$\delta u(x)$, b) ratio of fragmentation functions 
$H_1^{\perp (1) u} (z) / D_1^u (z)$.
The star in a) shows the normalization point.
The error bars correspond to an assumed data sample of 7 million DIS events.
The hatched bands show projected systematic uncertainties
due to the normalization and the $u$-quark dominance assumptions.
\label{fig:du_p}}
\vspace*{-0.7cm}
\end{figure}
The factorized form of Eq.~(\ref{uasymmetry}) with respect to
$x$ and $z$ allows the simultaneous reconstruction
of the shape for both unknown functions $\delta u(x)$ and
$H_1^{\perp(1)u}(z) / D^u_1(z)$ if measurements of the asymmetry
are done in $(x, z)$ 
bins. The relative normalization cannot be fixed without further 
assumption. Since differences between $\delta q(x)$ and $\Delta q(x)$ are 
expected to be smallest in the region of intermediate and large values of 
$x$~\cite{KNO}, $ \delta q (x_0) = \Delta q(x_0) $ at $x_0=0.25$ has been
assumed to resolve the normalization ambiguity.

Based upon a statistics of 7 million reconstructed DIS (Monte Carlo) events,
corresponding to about 150 nb$^{-1}$,
the projections for a measurement of $\delta u(x)$ and 
$H_1^{\perp(1)u}(z) / D^u_1(z)$ through $\pi^+$ production at HERMES have 
been calculated~\cite{KNO} for a mean target polarization of $P_T = 75 \%$,
as shown in Fig.~\ref{fig:du_p}.

\section{Improved Measurement of Helicity Distributions}
Access to quark helicity distributions requires both beam and target to be
longitudinally polarized. The measurement of double-spin cross section 
asymmetries in inclusive DIS directly allows for the extraction of $g_1$,
the longitudinal spin structure function which represents a given combination
of quark helicity distributions (for a recent review 
see e.g. Ref.~\cite{FilipJi}). To disentangle the helicity distributions of
individual quark flavours more observables are required. Of particular interest
are independent measurements of double-spin asymmetries for the production of
different hadrons $h$ in SIDIS:
\be
A_1^h(x,z)=
\sum_q\left[{\frac{e_q^2~q(x)\cdot D_1^{q,h}(z)}
{\sum_{q'}e_{q'}^2~q'(x)\cdot D_1^{q',h}(z)}}\right]
\cdot\frac{\Delta q(x)}{q(x)}
\ee
The quantity enclosed in brackets is the purity ${\cal P}_q^h(x,z)$, representing 
the probability that a quark $q(x)$ was struck when a hadron $h$ is detected. 
The underlying quark-hadron transition is described by the fragmentation function 
$D_1^{q,h}(z)$. It is assumed that the fragmentation of longitudinally polarized 
quarks is spin independent. Hence purities can be obtained from Monte Carlo 
calculations based on the same assumption. 
\begin{figure}[htb]
\centering
\vspace{-0.2cm}
\epsfxsize=6.7cm
\epsfbox{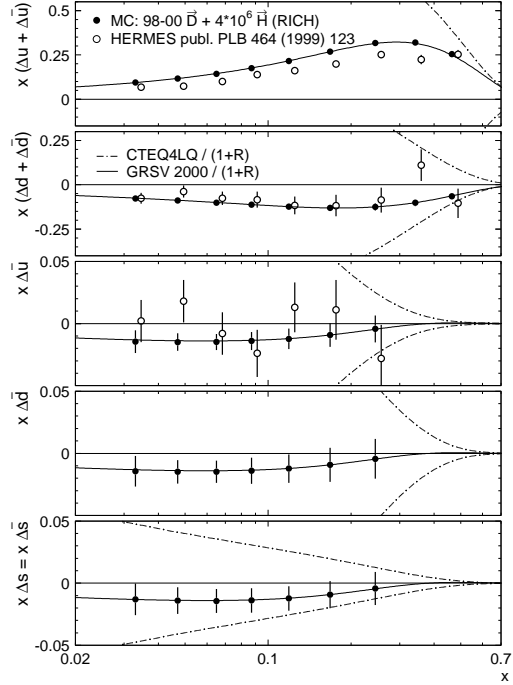} 
\vspace{-0.3cm}
\caption{Projected statistical accuracies for individual quark flavour helicity
distributions on the basis of the full 1996-2000 HERMES statistics 
plus one additional year of running with a longitudinally polarized proton target
(closed circles). The published HERMES results, derived from the 1995-1997 data,
are also shown (open circles).
\label{fig:delta_q}}
\vspace{-0.8cm}
\end{figure}

Combining DIS and SIDIS asymmetries for different targets, a fitting procedure
is used to determine the polarizations $\frac{\Delta q}{q}$ of individual 
quark flavours (for more details of the formalism see Ref.~\cite{Marc}). Quark 
flavours with low relative abundance, as all sea contributions, require a relatively 
large amount of statistics for the individual determination of their helicity 
distribution.

The most recent HERMES results are discussed in Ref.~\cite{Marc}. 
This analysis is preliminary and the statistics, on which it is based 
upon (it includes only part of the deuterium data already collected), are still 
insufficient for a precise determination of individual sea quark helicity 
distributions. Even when the total available statistics will have been analyzed, 
it will be difficult to draw unambigous conclusions on the polarization of 
the strange sea, $\Delta s$. Also, a discrimination between different theoretical 
models for the flavour asymmetry of the polarized light sea,
$\Delta \bar{u} - \Delta \bar{d}$, will remain difficult.

A considerable improvement could be reached by running HERMES another year with 
a longitudinally polarized proton target. The then existing total data set, a 
mixture of real and MC data, was fitted as mentioned before using 0.45 as mean
value for the combined beam and target polarization~\cite{MarcJuergen}.
As can be seen from Fig.~\ref{fig:delta_q} (closed symbols), the expected statistical 
accuracy looks very promising.

\section{Measurement of Exclusive Reactions to Access GPDs}
\begin{figure}[htb]
\centering
\vspace*{-2.4cm}
\epsfxsize=6.5cm
\epsfbox{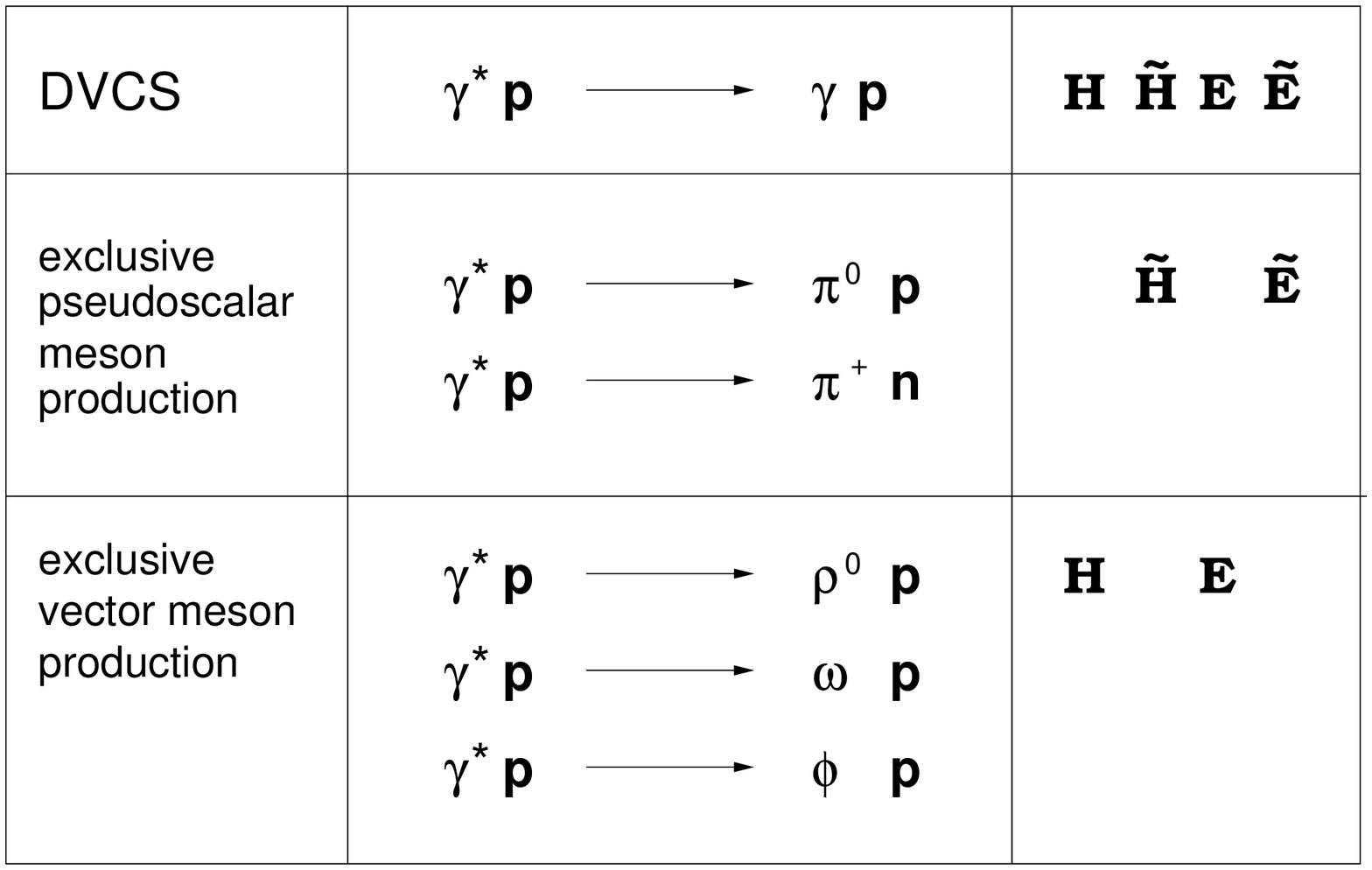} 
\vspace*{-2.4cm}
\caption{Hard exclusive processes accessible through HERMES in conjunction
with the involved Generalized Parton Distributions.
\label{fig:GPDreactions}}
\vspace*{-0.4cm}
\end{figure}

In the recent past a strong interest has emerged in Generalized Parton 
Distributions~\cite{GPDgen}, a set of non-perturbative process-independent 
distribution functions describing simultaneously different hard processes
from inclusive to hard exclusive scattering. Due to the good capabilities 
of the HERMES spectrometer outgoing photons as well as pseudo-scalar 
and vector mesons can be detected in a large variety of (quasi-) exclusive 
reactions. The relevant exclusive channels are illustrated in 
Fig.~\ref{fig:GPDreactions} in conjunction with the quark GPDs (one per flavour) 
that are in principle accessible.

\begin{figure}[htb]
\vspace*{-0.6cm}
\epsfxsize=5cm
\hspace{1cm}
\epsfbox{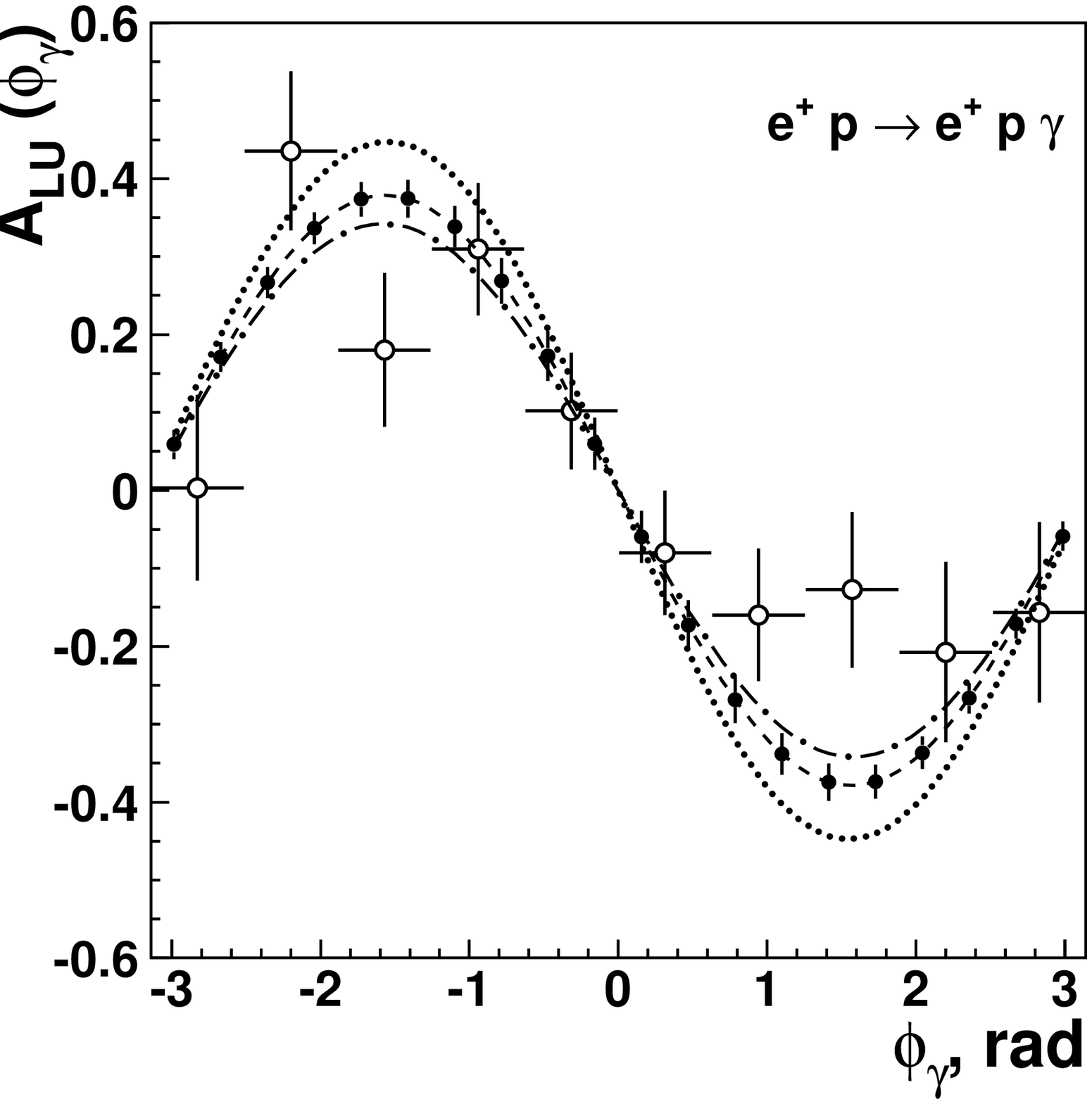}
\vspace*{-0.3cm}
\caption{As previous figure, but for the lepton helicity asymmetry $A_{LU}$ 
(polarized beam, unpolarized target). Open circles in left panel are existing 
HERMES data (see text).
\label{fig:helicity_asy}}
\vspace*{-6.05cm}
\hspace{6cm}
\epsfxsize=5cm
\epsfbox{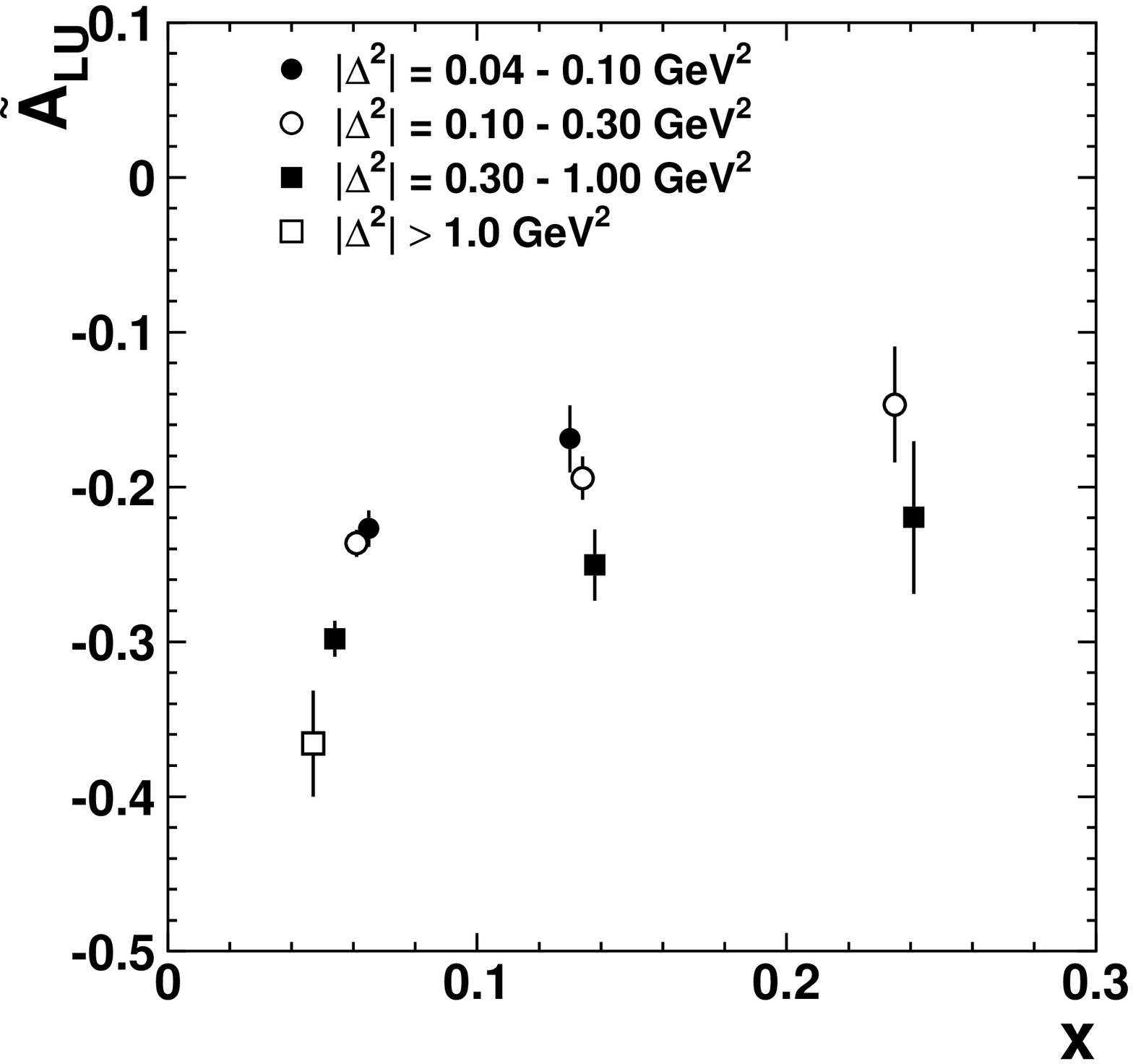} 
\vspace*{0.2cm}
\end{figure}

Deeply-Virtual Compton Scattering (DVCS) offers the most direct connection between 
theory and experiment, since no fragmentation is involved. At HERMES energies the
experimentally indistinguishable Bethe-Heitler (BH) process has a comparable
cross section and hence allows the study of DVCS through their 
interference. 
From 1996/97 HERMES data on quasi-exclusive photon production 
a sine-shaped single beam-spin asymmetry with a magnitude of about 
20\% has been extracted very recently~\cite{Jochen} (cf. 
fig.~\ref{fig:helicity_asy} left). 
It shows the beam-helicity dependence expected from BH-DVCS interference.

Having both beam and target unpolarized, the DVCS lepton charge 
asymmetry~\cite{Bel}
$A_{ch} \sim d \sigma(e^{+} p) - d \sigma(e^{-} p)$ exhibits a
cosine behaviour in $\phi_{\gamma}$, the azimuthal orientation of the 
$\gamma-\gamma^*$-plane w.r.t. the lepton plane,
\begin{eqnarray}
A_{ch}
&\sim& \cos(\phi_\gamma) \times 
{\rm Re} \left\{F_1 {\cal H}_1
+ \frac{x}{2 - x} (F_1+F_2) \widetilde{\cal H}_1
- \frac{\Delta^2}{4 M^2} F_2 {\cal E}_1\right\} ,
\end{eqnarray}
allowing access to the real parts of the DVCS amplitudes ${\cal H}_1$ 
and $\widetilde{\cal H}_1$, while ${\cal E}_1$ is suppressed. 
Bjorken-$x$ and $t$-channel momentum-transfer are denoted by $x$ and $\Delta^2$,
mass and form factors of the proton by $M$ and $F_i$, respectively.

The helicity of the polarized positron (or electron) beam of HERA is re-
\newpage
\noindent
gularly changed to minimize systematic uncertainties. In conjunction with an 
unpolarized target the DVCS lepton helicity asymmetry~\cite{Bel}
$A_{LU} \sim d\sigma(\overrightarrow{e^+}p) - d\sigma(\overleftarrow{e^+}p)$
becomes accessible. It exhibits a $\sin{\phi_{\gamma}}$-
behaviour,
\begin{eqnarray}
A_{LU}
&\sim& \sin(\phi_\gamma) \times
{\rm Im} \Bigg\{ F_1 {\cal H}_1
+ \frac{x}{2 - x} (F_1 + F_2) \widetilde{\cal H}_1
- \frac{\Delta^2}{4 M^2} F_2 {\cal E}_1 \Bigg\} ,
\end{eqnarray}
allowing access to the imaginary parts of ${\cal H}_1$ and $\widetilde{\cal H}_1$.

The real and imaginary parts of the DVCS amplitudes
${\cal H}_1, \widetilde{\cal H}_1, {\cal E}_1, \widetilde{\cal E}_1$ are convolutions
of known coefficient functions with the 4 different quark GPDs $H, \widetilde{H}, E$ 
and $\widetilde{E}$ (see e.g. Ref.~\cite{Bel}) shown in Fig.~\ref{fig:GPDreactions}. 
Their extraction from a set of asymmetries measured in different reactions is a 
complex task for which no proven recipe exists as of today.
\vspace*{-0.6cm}
\begin{figure}[htb]
\epsfxsize=5cm
\hspace{1cm}
\epsfbox{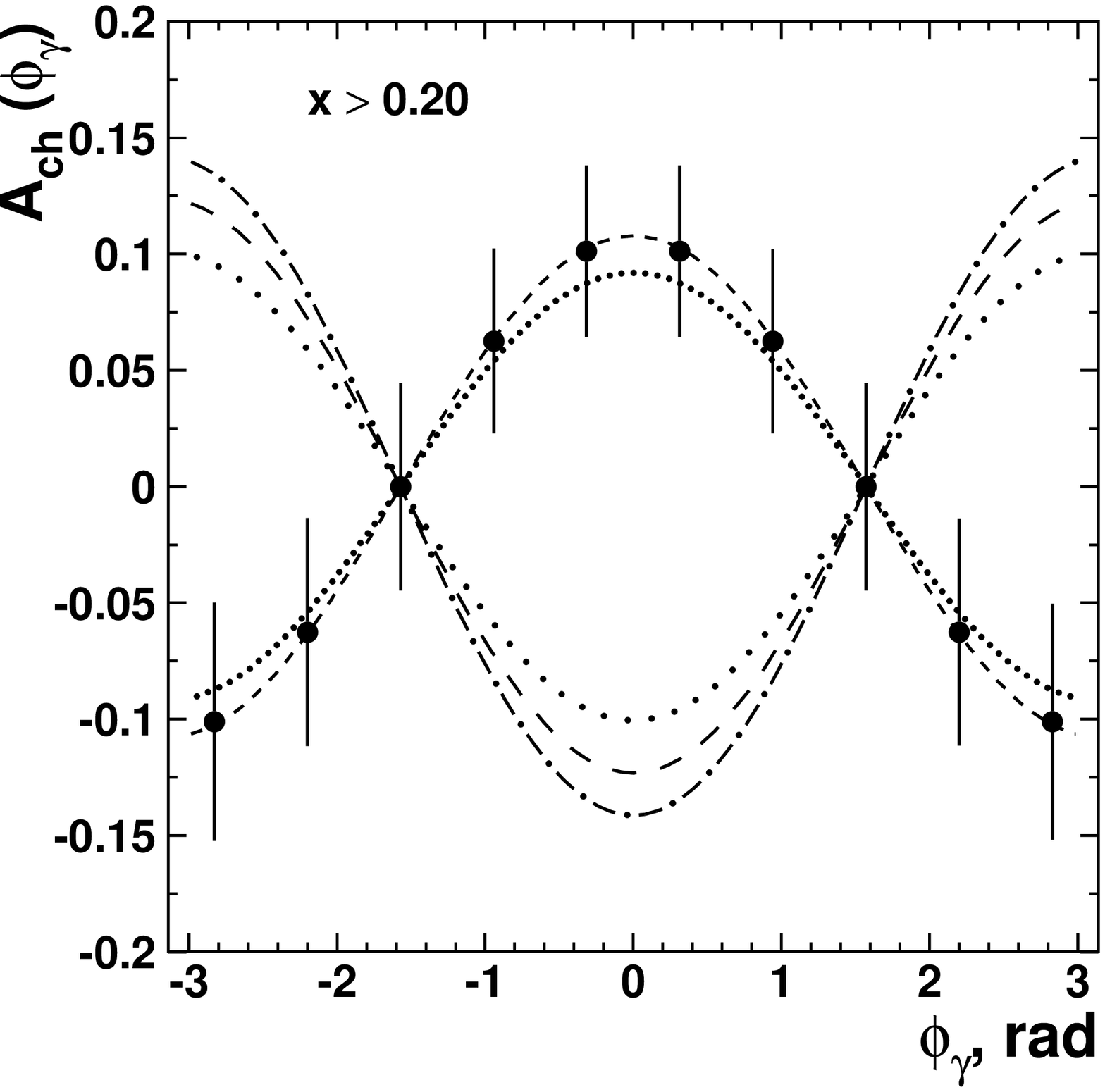}
\vspace*{-0.4cm}
\caption{Projected statistical accuracies for a HERMES measurement of the lepton charge 
asymmetry in DVCS (unpolarized beam, unpolarized target). Left panel: $\phi_{\gamma}$-
dependence of $A_{ch}$. The curves correspond to different choices of GPDs.
Right panel: $x$- and $\Delta^2$-dependence of $\tilde{A}_{ch}$,
see text. The error bars are based on an assumed data set of 2 fb$^{-1}$.
\label{fig:charge_asy}}
\vspace*{-6.85cm}
\hspace{6cm}
\epsfxsize=5cm
\epsfbox{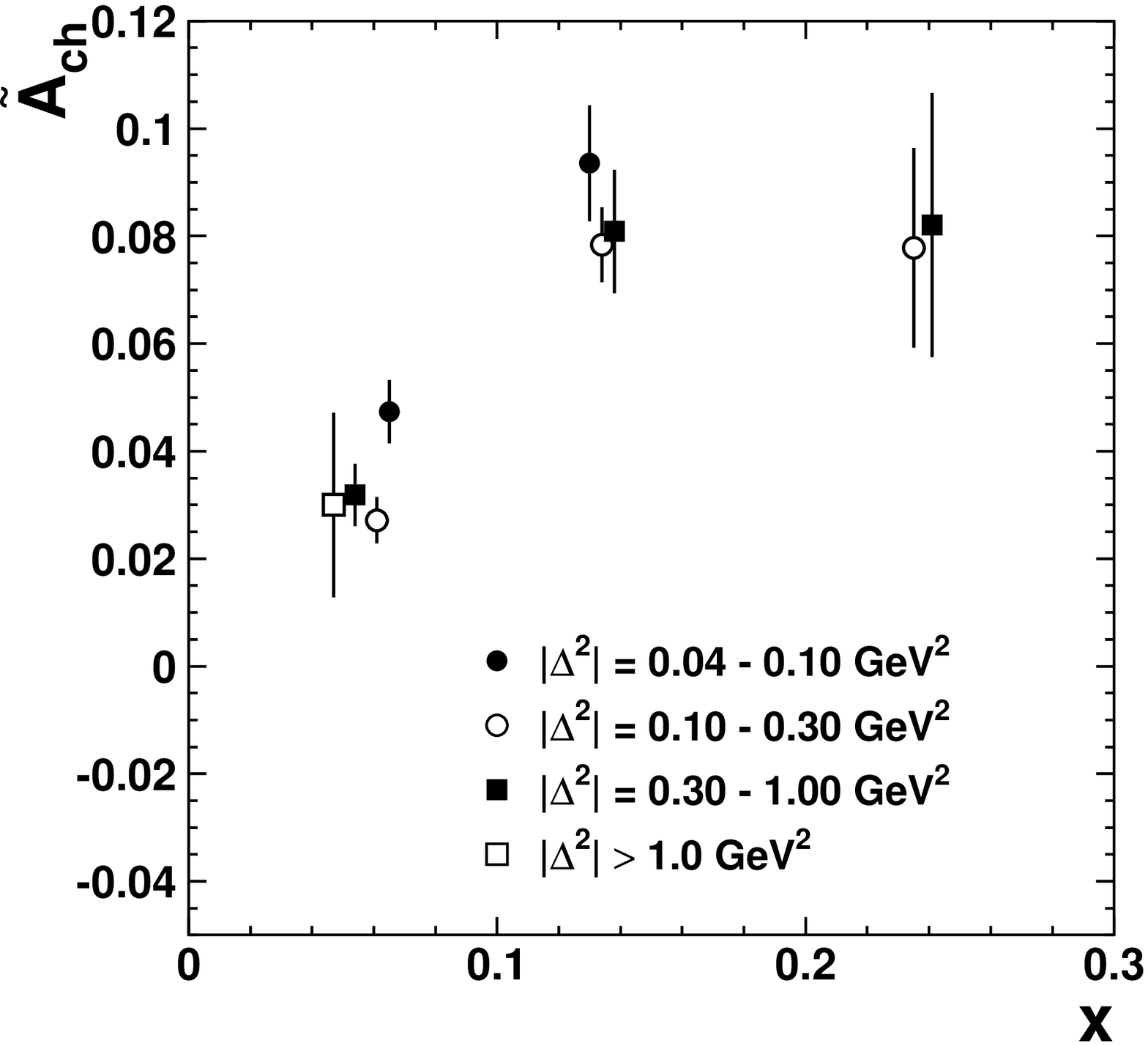} 
\vspace*{1cm}
\end{figure}

At present the missing mass resolution of the HERMES spectrometer is preventing 
an experimental exclusion of low-lying excited nucleon states ($\Delta$ isobars). 
Plans exist to add a Recoil Detector to the experiment 
to clearly identify exclusive 
events. High statistics data sets on exclusive photon, pion and meson production 
are expected from future running with a high density unpolarized target allowing
for an annual integrated luminosity of 2 fb$^{-1}$.

Based upon this experimental scenario the leading twist-2 formalism has been used 
to determine prospects for future DVCS measurements~\cite{KorNowDVCS}
taking into account the acceptance for all involved particles. 
%
The expected statistical accuracies are shown in 
Fig.~\ref{fig:charge_asy} for the lepton charge asymmetry $A_{ch}$, and in
Fig.~\ref{fig:helicity_asy} for the lepton helicity asymmetry $A_{LU}$. 
The left panels show the azimuthal dependence. The right ones display the
anticipated dependence on $x$ and $\Delta^2$, presented for $\tilde{A}_{LU}$,
the difference between asymmetry integrals over appropriate 
$\phi_{\gamma}$-hemispheres. Clearly, the ansatz for the GPDs directly 
influences the size of the asymmetry~\cite{GPDgen}. The $\Delta^2$-dependence 
will become experimentally accessible at HERMES through the new Recoil Detector. \\

%
\noindent {\bf 5 Conclusions}
\vspace*{1mm} \\
The main physics objectives for the next 5 years of HERMES running have been 
sketched: Measurements of SIDIS using transversely and longitudinally
polarized proton targets, and of exclusive reactions using unpolarized targets.
Information on several hitherto unknown quantities will be provided to shed 
light on different new aspects of the nucleon (spin) structure.



\end{document}